%% file: main.tex
\documentclass[runningheads]{llncs}
\usepackage[T1]{fontenc}
\usepackage{newtxtext}
\usepackage[varvw]{newtxmath}
\usepackage{graphicx}
\usepackage{orcidlink}

\usepackage{algorithm}
\usepackage[noend]{algpseudocode}
\usepackage{listings}
\usepackage{xcolor}
\usepackage{booktabs}
\usepackage{bbding}
\usepackage{longtable}

\usepackage{tikz} %  huub: added for the images
\usetikzlibrary{positioning,arrows.meta,shapes.geometric,calc,fit,shapes.symbols}

\begin{document}

\title{LLM-Assisted Verification in Dafny}

\author{
  Wieger Wesselink$^{\mbox{\Envelope}}$\orcidlink{0009-0001-6746-5115} \and
  Kees Huizing\orcidlink{0000-0002-4713-7253} \and
  Huub van de Wetering\orcidlink{0000-0002-0517-1322}
}
\authorrunning{W. Wesselink et al.}
\institute{Eindhoven University of Technology, The Netherlands \\
  \email{\{j.w.wesselink,h.v.d.wetering,c.huizing\}@tue.nl}
}

\maketitle

\begin{abstract}
Formal verification of complex algorithms requires expertise in program specification, invariant construction, mathematical proof, and the management of automated theorem provers. Although large language models (LLMs) can generate program code and proof fragments, reliably developing complete machine-checked verifications remains challenging: it requires an iterative process of constructing and validating specifications and invariants, building proofs, diagnosing solver failures, and optimizing verification performance. This paper presents a systematic ten-step workflow for LLM-assisted verification of scientific algorithms using the Dafny verification system. The workflow integrates algorithm formalization, contract development, empirical invariant validation, proof construction, bottleneck profiling, and resource-driven proof optimization. We applied the workflow using Antrhopic's Claude Opus 5 and Opus 5.5 models, combining LLM-driven development with automated testing and verification feedback, while retaining human oversight for correctness-critical decisions and strategic interventions. We evaluate the approach on 37 complex algorithms from graph theory and labelled transition systems. Notably, we verified the Hopfcroft--Tarjan triconnectivity test, demonstrating that even highly challenging algorithmic verification tasks can be carried out through a structured LLM-assisted workflow. Many of our machine-checked verifications were obtained within hours to a few days of interactive development. Our results suggest that LLM-assisted workflows can shift the role of the human verifier from detailed proof construction toward supervision, specification review, and strategic guidance.
\end{abstract}

%%%%%%%%%%%%%%%%%%%%%%%%%%%%%%%%%%%%%%%%%%%%%%%%%%%%%%%%%

\section{Introduction}
\label{sec:introduction}

Formal verification establishes that a program satisfies a
mathematical specification. Unlike testing, it provides a proof
of correctness rather than checking a finite set of executions.
This is particularly valuable for complex algorithms, whose
correctness often depends on subtle properties of their data
structures and control flow. However, developing machine-checked
proofs for such algorithms remains difficult. Even when the
underlying mathematical argument is known, expressing it in a
form that an automated verifier can establish may require
substantial proof engineering.

Dafny is a programming language and verification system that
combines program development with formal specifications and
proofs~\cite{Leino10}. It supports preconditions, postconditions,
loop invariants, and termination measures, and translates the
resulting verification conditions into queries for the Z3 SMT solver.
Although many proof steps are automated, the
developer must still provide suitable invariants, lemmas, and
intermediate assertions, and organise the proof so that the
solver can handle it. Large verification contexts, quantified
properties, and interactions between assertions can make
verification difficult even when the required mathematical
facts are available.

Large language models (LLMs) offer a way to reduce manual
effort involved in this process. They can develop program code,
propose invariants and lemmas, construct proofs, and interpret
verification errors. However, generating proof steps is only
part of the task. A complete verification effort also involves
formalising the algorithm, developing and testing an
implementation, validating invariants, diagnosing solver
problems, and ensuring that the complete program verifies.
These activities require an iterative process in which the
LLM uses feedback from testing and verification to guide
further development.

Recent work on human-directed agentic engineering has shown
the value of combining LLMs with tools, structured workflows,
checkpoints, and human oversight~\cite{Kurtzer2026Ginux}.
We propose a similar approach to formal verification. Rather
than focusing on a particular invariant-generation or
proof-synthesis technique, we treat LLM-assisted verification
as a workflow design problem. The aim is to combine LLMs,
Dafny, empirical testing, solver measurements, and human
review into a systematic process for constructing
machine-checked proofs.

We use Claude Code, Anthropic's agentic coding assistant, to construct and verify Dafny implementations of published algorithms. In our workflow, Claude Code is used to develop specifications, generate implementations, design loop invariants, construct proofs, and interact with the Dafny verifier. We refer to Claude Code in this role as the Verification Construction Agent (VCA). The VCA operates within a workflow that provides it with tools, instructions, and access to a persistent repository. Its work is supervised by a human, who reviews the verification target and provides strategic direction during proof development.

This paper presents a ten-step workflow for LLM-assisted
verification of algorithms in Dafny. Starting from algorithm
descriptions in the literature, the workflow covers
formalisation, contract development, implementation and testing,
invariant construction and empirical validation, proof
construction, and verification optimisation. It combines
feedback from empirical invariant checks with measurements
of solver resource consumption to guide proof development.
The process is iterative: verification failures may require
changes to invariants, proof structure, or even the chosen
strategy. Human supervision remains important, particularly
when the workflow needs to change direction.

We evaluate the workflow on more than thirty algorithms,
primarily from graph theory and labelled transition systems
(LTSs). The collection includes challenging targets such as
Edmonds' blossom algorithm, Lengauer--Tarjan dominators, and
the Hopcroft--Tarjan triconnectivity test. The results show
that the workflow can produce substantial machine-checked
verifications across different algorithm families. They
also show that successful verification requires more than
discharging individual proof obligations: the complete
program must also verify under Dafny's native configuration.
This distinction is important because proof structure and
solver behaviour can cause a program to fail even when
its obligations verify successfully in isolation.

The main contributions of this paper are the following.
\begin{itemize}
    \item We present a ten-step workflow for developing
    machine-checked Dafny verifications from algorithm
    descriptions in the scientific literature.

    \item We integrate empirical invariant validation,
    including truth checks, counterexamples to induction,
    and drop tests, into proof development.

    \item We present a resource-guided approach to proof
    engineering, using verifier measurements to identify
    expensive obligations and guide proof restructuring
    and optimisation.

    \item We evaluate the workflow on more than thirty
    algorithms, including challenging graph and LTS
    algorithms, and, in particular, report a machine-checked verification
    of the Hopcroft--Tarjan triconnectivity test.
\end{itemize}

The remainder of the paper is organised as follows.
Section~\ref{sec:scope} defines the verification challenge
and scope. Section~\ref{sec:related-work} lists the related work.
Section~\ref{sec:workflow} presents the ten-step
workflow. Section~\ref{sec:tools} describes the tools
and Section~\ref{sec:key-techniques} the proof-engineering techniques. 
Section~\ref{sec:llm} shows cross-verification knowledge reuse. 
Section~\ref{sec:evaluation}
presents the evaluation,
Section~\ref{sec:results} discusses the results, and
Section~\ref{sec:conclusions} concludes the paper and
outlines future work.

%%%%%%%%%%%%%%%%%%%%%%%%%%%%%%%%%%%%%%%%%%%%%%%%%%%%%%%%%

\section{Verification Challenge and Scope}
\label{sec:scope}

We focus on the functional verification of algorithms described
in the scientific literature, rather than arbitrary existing
programs. Starting from published algorithm descriptions,
typically given as pseudocode, we develop Dafny implementations
and establish their correctness with respect to contracts based
on standard definitions. This distinguishes our
approach from methods that start with existing implementations
and mainly generate missing annotations or proofs. Low-level
implementation details, memory management, and language-specific
optimisations are outside our scope. We also do not assess the
pseudocode notation itself or claim that formal verification
removes all ambiguity in the interpretation of an algorithm.

We use the term \emph{complex algorithm} for an algorithm whose
structure and correctness properties make formal verification
substantially more difficult than for simple textbook examples.
The algorithms considered include graph and labelled-transition-
system algorithms with nested loops, large loop bodies, mutable
data structures, and iterative or recursive control flow. Their
correctness often depends on invariants and global properties
involving vertices, edges, paths, or other structures. Examples
include connectivity, reachability, minimality, and structural
characterisations. Establishing termination may also require
non-trivial progress measures.

A major challenge in verifying these algorithms is
\emph{proof engineering}. The reasoning must be
expressed through invariants, lemmas, and assertions in a form
that a proof checker can handle. This requires careful control
of proof contexts, decomposition of verification conditions,
and structuring of assertions. Solver performance can be
sensitive to seemingly small changes in proof structure and
assertion batching. As a result, even a mathematically
straightforward proof can be difficult to verify automatically,
and proof-engineering effort does not necessarily grow
predictably with algorithmic complexity.

Our work focuses on Dafny and its underlying verification
toolchain. The workflow is therefore shaped by Dafny's
verification model, Z3's capabilities, and the practical
challenges of verifying large programs reliably. 
%Applying similar ideas to other verification systems is left for future work.

%%%%%%%%%%%%%%%%%%%%%%%%%%%%%%%%%%%%%%%%%%%%%%%%%%%%%%%%%

\section{Related Work}
\label{sec:related-work}

Dafny has been used to verify nontrivial algorithms and programs, including solutions to verification competitions, case studies, and modular proof development~\cite{farrell2021using,faria2023case,ho2024incremental}. Related work has also studied techniques for diagnosing difficult proof obligations~\cite{gopinathan2025accelerating}. Large collections of verified algorithms have been developed in interactive theorem provers such as Isabelle/HOL~\cite{NipkowPW02}, Coq~\cite{BertotC04}, and Lean~\cite{Moura021}. While these systems generally require more explicit proof construction than Dafny, they also provide automation. Studies of verification-aware programming have highlighted the effort and expertise needed to develop specifications and proofs~\cite{mugnier2025impact,oliveira2025challenges}.

Recent work has explored the use of large language models (LLMs) to assist Dafny verification. Approaches include lemma generation, annotation synthesis, and loop invariant generation~\cite{silva2024leveraging,poesia2024dafnyannotator,faria2025invariants}. Further work addresses the generation of formal specifications and verification annotations~\cite{DBLP:journals/corr/abs-2601-12845} and the repair of arithmetic errors in existing Dafny programs~\cite{DBLP:conf/sefm/WuMA25}. DafnyPro~\cite{banerjee2026dafnypro} improves generated annotations using techniques such as invariant pruning and retrieval of successful proof strategies. ATLAS~\cite{Baksys25a} explores the large-scale synthesis of complete verified programs, while miniF2F-Dafny~\cite{pmlr-v306-baksys26a} applies LLMs to mathematical theorem proving with Dafny. These approaches address important parts of the verification process, but focus on specific tasks or different synthesis goals.

LLMs have also been studied for formal specification generation and verified program synthesis beyond Dafny. SpecGen~\cite{DBLP:conf/icse/MaL0XB25} and Le-Cong et al.~\cite{DBLP:conf/acl/Le-CongLM25} investigate the generation and inference of formal specifications. The Vericoding benchmark~\cite{DBLP:journals/tmlr/ChiosaBELAKSBBDZTT26} evaluates the generation of formally verified programs. These works provide relevant approaches and evaluation settings, but do not focus on the complete development of algorithms from descriptions in the scientific literature.

Our work addresses this broader development process. Starting from an algorithm description, we use LLMs to develop its specification and implementation, discover and validate invariants, construct proofs, and refine the program through verifier feedback and performance measurements. Unlike approaches that focus on individual verification tasks or assume an existing implementation, our workflow covers the development process from algorithm description to verified implementation. It combines LLM assistance with empirical validation, resource-guided proof development, and human supervision.

%%%%%%%%%%%%%%%%%%%%%%%%%%%%%%%%%%%%%%%%%%%%%%%%%%%%%%%%%

\begin{figure}[tbh]
  \centering
  \resizebox{1.0\linewidth}{!}{\input{workflow}}
  \caption{Conceptual ten-step workflow for LLM-assisted verification. Starting from an article Steps~1--2 establish a contract plus pseudocode 
that after human approval becomes \emph{locked}. In Steps~3-4 Dafny and Python implementations are generated and the contract is tested with randomized inputs.  Steps~5--6 create a proof plan consisting of empirically tested loop invariants and termination measures. Step~7 converts this plan into a draft Dafny proof. 
Steps~8--9 then iteratively verify and optimize the proof using subagents for solving individual issues. All steps in this loop are under human supervision.
Finally, step~10 collects the experiences gained during the process. They are stored in the repository (dashed box) for usage in the current or future verification efforts.
\label{fig:workflow}}
\end{figure}
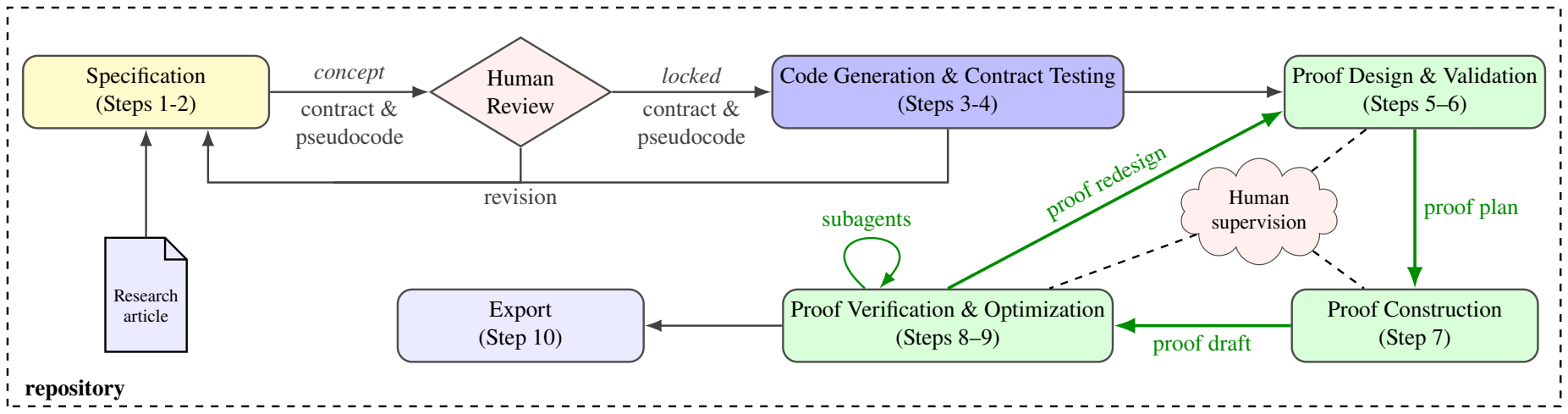

\begin{table}[ht]
\centering
\caption{The ten workflow steps and Claude's contribution.}
\label{tab:steps}
\small
\footnotesize
\begin{tabular}{@{}r l p{33mm} @{\hspace{2mm}} p{63mm}@{}}
  \toprule
  \# & Step & Produces & Claude's action \\
  \midrule
  1 & \texttt{spec} & Pseudocode in \LaTeX{} &
    Transcribes the publication; identifies ambiguities, and proposes resolutions \\

  2 & \texttt{contract} & Formal pre- and post\-conditions; contract lock &
    Formalizes the intended functional behavior; adds missing domain definitions to the Dafny and Python libraries \\

  3 & \texttt{generate} & Python and Dafny implementations &
    Runs the generators and checks that the generated Dafny code is well-formed and type-correct \\

  4 & \texttt{test} & Randomized Python tests &
    Checks  pre- and postconditions using randomized tests in Python \\

  5 & \texttt{invariants} & Loop invariants and termination measures &
    Writes invariants and termination measures; establishes the initialization, preservation, and exit obligations for each loop \\

  6 & \texttt{conjectures} & Empirically tested invariants; lemma plan &
    Writes an instrumented Python copy; empirically tests each invariant; plans the lemmas \\

  7 & \texttt{complete} & Unverified proof draft &
    Writes lemmas and ghost code without running Dafny; reuses proofs from similar algorithms \\

  8 & \texttt{verify} & Individually verified proof obligations &
    Runs a resource-restricted subagent loop to verify each obligation in isolation\\ %: diagnose, repair, and rerun \\

  9 & \texttt{optimize} & Proof verified as a whole &
    Runs a subagent loop until the entire file verifies in Dafny; does an audit for implicit assumptions\\%: measure, restructure, and rerun; checks that every assumption is justified \\

  10 & \texttt{export} & Session transcripts &
    Logs transcripts to the repository \\
  \bottomrule
\end{tabular}
\end{table}

\section{The Workflow}
\label{sec:workflow}

We have defined an interactive, ten-step workflow executed by Claude Code, which serves as the Verification Construction Agent (VCA) under human supervision. The VCA uses a repository to maintain persistent information about the verification process and its results. It also has access to supporting tools, for example, to identify verification bottlenecks. Throughout the workflow, guardrails constrain the VCA's actions, for example, by preventing unauthorized changes to the algorithm and limiting resource usage.

The workflow organizes the construction of machine-checked Dafny verifications into ten steps, see  Figure~\ref{fig:workflow}. It starts from an algorithm described in the literature and produces a formal specification, executable implementations, invariants, proofs, and verification artefacts. The steps have defined inputs, outputs, and completion criteria, summarized in Table~\ref{tab:steps}. The steps can be divided into three main phases: a specification phase that establishes a contract plus pseudocode, followed by a code generation and contract testing phase, and the main proof-development phase (the green loop in Figure~\ref{fig:workflow}), where the human works with the VCA.

\paragraph{Establishing the verification target (Steps~1--2).}
In Step~1, the agent reconstructs the algorithm from its publication into the workflow's pseudocode representation. It also identifies ambiguities and proposes interpretations where necessary. In Step~2, it formalizes the intended behavior as a contract consisting of preconditions and postconditions, adding any required definitions to the Python and Dafny domain libraries. The human reviews the pseudocode, the recorded interpretations, and the contract. Once approved, the contract is locked and cannot be changed without explicit human approval.

\paragraph{Generating and testing the implementations (Steps~3--4).}
In Step~3, the agent uses tooling to generate Python and Dafny implementations from the pseudocode and contract. The generated Dafny code is checked for well-formedness and type correctness. In Step~4, the Python implementation is tested using randomized inputs. The tests check the preconditions before execution and the postconditions after execution. The resulting implementations and test results provide the starting point for proof development.

\paragraph{Developing the proof plan (Steps~5--6).} In Step~5, the agent constructs loop invariants and termination measures. It identifies the properties needed to establish initialization, preservation, and the loop exit conditions. In Step~6, candidate invariants are tested using an instrumented Python implementation. The workflow checks whether invariant clauses hold in reachable states, whether they are preserved by program transitions, and whether individual clauses are needed for inductiveness.  Based on these results, the agent refines the invariants and develops a plan for the supporting lemmas.

% \paragraph{Constructing and verifying the proof (Steps~7--9).}
% In Step~7, the LLM translates the proof plan into a Dafny proof draft, consisting of lemmas and ghost code. It structures the proof around individual state transitions and reuses existing proofs where applicable. In Step~8, the proof obligations are verified individually. Failed obligations are diagnosed and revised in resource-restricted subagent loops until they verify. 
% In Step~9, the complete Dafny file is verified under the normal verification configuration. Verification performance is measured, and bottlenecks are identified using a dedicated support tool. This tool is an important part of the workflow: it helps locate proof obligations or parts of the implementation that dominate verification time, so that the VCA can focus its optimization efforts. The VCA then restructures the proof where necessary and repeats the verification and measurement process. This continues until the entire file verifies within the intended resource limits. The final proof is also checked for unjustified assumptions.

\paragraph{Constructing and verifying the proof (Steps~7--9).}
In Step~7, the VCA translates the proof plan into a Dafny proof draft containing lemmas and ghost code, reusing existing proofs where applicable. Steps~8 and~9 involve substantial proof engineering. In Step~8, individual proof obligations are verified, and failures are diagnosed and addressed in resource-restricted subagent loops. This may require correcting claims, adding missing proof facts, or restructuring proofs to reduce solver context. In Step~9, the complete file is verified and its performance measured. A dedicated tool identifies bottlenecks, guiding further proof engineering through techniques such as lemma decomposition, controlling solver context, and checking assertions in isolation. Both steps involve iterative diagnosis and refinement, continuing until the complete file verifies within the intended resource limits. The final proof is also audited using Dafny for unjustified assumptions. Section~\ref{sec:key-techniques} describes the techniques in detail. Changes made during proof development or verification may also require revisiting Steps~5--6 to revise the invariants or termination measures.

\paragraph{Recording the results (Step~10).}
In the final step, the workflow stores the artefacts and results of the verification into the repository, including specifications, contracts, implementations, proofs, and session records.

\paragraph{Human supervision} The human supervises the proof-development process at both the strategic and technical levels. At the strategic level, the human monitors progress and verification results and may intervene to change the approach when the current strategy is ineffective or inefficient. For example, the human may ask the VCA to investigate several unfinished proof obligations together, identify a common set of missing facts, and only then return to revising the invariants or contracts. The VCA translates such directions into concrete verification tasks and carries them out. The human may also contribute ideas about how to proceed with a proof or suggest a proof technique. Thus, the human provides oversight and, when useful, strategic or technical guidance, while the VCA performs the detailed analysis and proof development.

\paragraph{Guardrails}
Guardrails are static checks on a Dafny file, enforced by tools. Their purpose is to make sure that a successful verification actually establishes the algorithm's stated correctness, and that edits have not weakened the specification or added unsound assumptions. They include checks that no changes are made in the generated code and its contract, checks that all added code is ghost code, and checks that no axioms, assume statements or unsound constructs are used. Some exceptions are allowed, e.g., adding \texttt{\{:isolate\_assertions\}} annotations.

% Every line of the generated snapshot must still be present in the edited file. The exceptions are a small allow-list: generator placeholders (requires true, ensures true, decreases *) and certain proof attributes on signatures.
% 2. Ghost declarations. Every hand-added function, predicate, method or lemma must be ghost, so proof code cannot change the behaviour of the executable algorithm.
% 3. Ghost statements. Every statement added to a method body must be ghost, for example ghost var rather than var.

% 7. Contract lock. The signatures, pre- and postconditions of the main methods, and every definition they depend on, must match the contract that was reviewed and locked in an earlier workflow step. This catches a strengthened precondition or a weakened definition, which all the other checks let through.

% 4. Postconditions present. Every method must have at least one ensures clause. Without one, a method verifies trivially.

% 5. No needless decreases *. A method may keep decreases * (no termination proof) only while a loop, recursive call or callee still needs it.
% 6. Registered axioms. Every assume and {:axiom} reachable from the file, including through included files, must be listed with a justification in a central exception registry.

% Guardrails restrict changes to reviewed specifications and generated code, and check that proof development does not introduce unjustified assumptions or unsound constructs.

\paragraph{Persistent storage}
The repository preserves the current verification state and provides access to previously developed proofs, domain definitions, and verification techniques, which are described in Section~\ref{sec:key-techniques}, while knowledge-reuse mechanisms are discussed in Section~\ref{sec:llm}.
 It also maintains a status report that summarizes the current state of the verification.
Importantly, the repository is not merely an archive of completed verifications. It supports learning and reuse across verification tasks. The workflow maintains a catalog of proof techniques and records lessons learned during individual verifications. These records can inform later tasks, for example by suggesting a proof strategy, identifying useful lemmas or domain definitions, or warning about approaches that previously caused difficulties. 
%The session transcripts provide evidence that this reuse occurs at several levels:
% This allows the VCA to reuse individual proof artefacts, apply established proof patterns, draw on domain-specific knowledge, and adapt its approach based on experience from earlier verification efforts.

% In summary, the workflow begins with human review and approval of the algorithm formalization and its contract. Once the verification target is established, the LLM can carry out the remaining implementation, proof-development, and verification tasks with a
% high degree of autonomy. In particular, the mathematical work in Steps~5--9 does not normally depend on human guidance, even for challenging algorithms or unfamiliar application domains. Guardrails
% constrain the process, and the repository preserves its state and evidence across sessions. The workflow thus combines human judgement at the point where the intended behavior is established with autonomous LLM-based construction of machine-checked proofs.
\paragraph{Summary} In summary, the workflow begins with human review and approval of the algorithm formalization and its contract. Once the verification target is established, the VCA carries out the implementation, proof-development, and verification tasks with a high degree of autonomy. The human supervises this work, provides strategic direction when appropriate, and may occasionally contribute proof ideas or suggest proof techniques. Support tools help the VCA test candidate invariants, verify proofs, and identify performance bottlenecks. The repository preserves the verification state and accumulates reusable proof techniques and lessons from previous tasks. The workflow thus combines human judgement, autonomous proof construction, tool-supported verification, and accumulated experience across verification efforts.

%%%%%%%%%%%%%%%%%%%%%%%%%%%%%%%%%%%%%%%%%%%%%%%%%%%%%%%%%%%%%%%%%%%%%%%%%%%

\section{Tools}
\label{sec:tools}

%\subsection{Tools}
The VCA uses specialized tools for code generation, integrity
checking, invariant testing, and solver diagnostics. Algorithms are
specified as LaTeX pseudocode with structured comments, from which corresponding Python and Dafny
implementations are generated.

\paragraph{Execution limits and integrity checks.}
Verification runs are executed under controlled resource limits (number
of parallel solver processes, memory per solver process, total memory,
and time per proof obligation); concurrent runs share a single memory
budget. Additional tools ensure that
generated programs and reviewed contracts are not modified
unintentionally. A contract lock records the reviewed contract together
with every definition it depends on, including library definitions,
since weakening a definition would weaken the contract without changing
its text; a changed contract is accepted only through an explicit step that
the user confirms. 
%Edits to generated code are insert-only, other
%changes to generated lines are restricted to explicit patches, proof
%additions must consist of ghost code, and uses of \texttt{assume},
%axioms, and related constructs are checked. These controls preserve the
%verification target throughout the development process.

\paragraph{Invariant checking.}
A dedicated checker tests candidate invariants before formal proof
construction begins. The checker executes the Python implementation on
random inputs and evaluates invariant clauses and callee postconditions
at the program points relevant to the proof. Three complementary
analyses are performed. A \emph{truth test} identifies clauses that are
false in reachable states. 
Consider a program fragment described by the Hoare triple \(\{P\}\ S\ \{Q\}\), where \(S\) is typically a loop body. A \emph{counterexample-to-induction (CTI) test} starts with a state \(s\) from a real execution, such that \(P(s)\) holds and executing \(S\) from \(s\) produces a state satisfying \(Q\). It then perturbs \(s\) to obtain a state \(s'\), retaining the perturbed state only if \(P(s')\) still holds. If executing \(S\) from \(s'\) produces a state that violates \(Q\), a counterexample to induction has been found. This demonstrates that \(P\) is too weak to establish \(Q\), even though both properties hold along all real executions.
Finally, \emph{drop tests} repeat the CTI test with
individual clauses removed to identify clauses that appear necessary for
inductiveness. Together, these analyses provide empirical evidence about
correctness, inductiveness, and necessity, helping to identify
incorrect, insufficient, or redundant clauses. They complement, but do
not replace, formal verification: the checker finds counterexamples but
proves nothing. %, and because callees are executed concretely, a fact
% missing from a callee's postcondition can go undetected.

\paragraph{Solver diagnostics and verification.}
The verification tools identify failing, timing-out, and expensive proof
obligations, reporting their declarations, relevant clauses, and solver
resource consumption. Additional diagnostics localize costly proof
blocks, inspect quantifier instantiations, and identify unsatisfied or
expensive callee preconditions. Verification is examined both in
isolation and under Dafny's default batched verification strategy. In
isolation, assertions are checked through separate solver queries; in
batched mode, multiple obligations share queries. Consequently, an
obligation that verifies in isolation may still fail or time out when
verified as part of a larger context. The workflow therefore performs
proof development and diagnosis primarily in isolated mode, followed by
measurement-driven optimization of the complete verification artefact
until it verifies under Dafny's native configuration. Verification runs
are performed using fixed solver settings so that the effects of
proof-structure changes can be compared reliably. The resulting
measurements and diagnostics guide the proof-engineering techniques
described below.

\section{ Proof-Engineering Techniques}
\label{sec:key-techniques}

A verification failure rarely reveals its underlying cause directly. A failed or timed-out obligation may indicate that a statement is false, that an invariant is too weak, that a proof fact is missing, or that the verifier is overwhelmed by the size of the proof context. The verification workflow comprises roughly 25 techniques for addressing such problems. This section presents the key techniques and explains how they address these different causes of verification failure. They act on intermediate claims, proof structure, or the context visible to the solver; the integrity checks described above ensure that none of them changes what is ultimately proven. The reported figures are based on interventions documented so far and should be regarded as indicative rather than exhaustive, since the analysis of the full corpus is still ongoing.

\paragraph{Check the claim before proving it.}
Some verification failures stem from statements that are false or fail the required induction step, which no proof restructuring can repair. The invariant checker is therefore applied twice: first to candidate invariants and intermediate claims before proving starts, and again to any failure that remains after all performance techniques have been tried. In the seven failures examined in this second way, the statements were false, and no solver tactic could have proven them.

\paragraph{Lemma decomposition.}
Large proofs are decomposed into small lemmas with simple, flat preconditions and narrowly defined responsibilities. Decomposition serves both correctness and performance: it localizes failures and allows expensive facts to be established once and reused. Its main performance benefit is that each lemma is proven in a small context of its own, making calls to it from a large method cheap. This is particularly useful for elementary facts about sets, cardinalities, arithmetic, and graph structure. Such a fact can be as expensive to prove as a hard obligation when embedded in a large method context, but nearly free when established once in a separate lemma. Lemma decomposition is used extensively throughout the verification: the results for 37 algorithms include a total of 2,351 lemmas.
%Decomposition is the background practice behind nearly every proof: the verified files contain 2,805 hand-written lemmas across 126 files.

\paragraph{Opaque specifications and proof in lemmas.}
In large methods, each verification condition exposes the solver to many contract facts, loop invariants, and quantified definitions at once, leading to excessive quantifier instantiation. To reduce this context, the predicates in the method's contract, its loop invariants, and the contracts of its callees are made opaque and bundled where useful. They are never revealed in the method body. Instead, every fact the method needs is established by a lemma that reveals only the definitions it requires. The method body therefore contains only code, bounds, and lemma calls. This technique works only when applied consistently: hiding some predicates while leaving others visible often failed, whereas the complete design succeeded. In Lengauer--Tarjan, a partially opaque main method still had 36 timeouts (2,098M resource units). The complete design eliminated all of them and reduced the cost to 144M. Among the documented applications, this was the most effective technique: all 11 applications ended in a verified method, including three of the hardest targets, after recorded alternatives had failed.

\paragraph{Assertion isolation.}
Dafny normally groups multiple assertions within a declaration into verification queries, allowing the solver to reason about them together. With assertion isolation, each assertion is checked separately, using the \texttt{\{:isolate\_assertions\}} annotation on the declaration. This can replace an expensive query with several queries that are individually cheaper to verify. For example, \texttt{bron\_kerbosch2} dropped from 113.2M resource units when assertions were batched to 5.2M when isolated. However, isolation can also duplicate work that the obligations would otherwise share, or fail outright: the main method of Bor\r{u}vka's algorithm ran out of memory under isolation until part of its context was hidden. The annotation must therefore be selected per declaration, based on measurement. It is the most frequently used technique among the interventions documented so far, and isolation annotations appear across many files.

\section{Knowledge Reuse}
\label{sec:llm}

% The infrastructure addresses two challenges in scaling LLM-assisted
% formal verification: managing verification efforts that exceed the
% limits of individual LLM sessions, and 
% preserving knowledge for reuse
% in later verifications. It combines a coordinator/subagent architecture
% for managing long-running tasks with mechanisms for reusing verified
% results and proof-engineering experience.

% \subsection{Coordinator and Subagent Architecture}
% \label{sec:subagents}

% Long-running verification tasks are managed by a coordinator that
% launches bounded proof-engineering subagents. Each subagent starts
% from the current repository state, works within a fixed resource
% budget, records its findings, and reports its outcome. When further
% work is needed, the coordinator launches a new subagent, which
% continues from the updated repository state rather than relying on
% conversational memory. This allows verification efforts involving
% hundreds of verification runs and many iterations of proof engineering
% to span multiple independent LLM sessions.

% The coordinator tracks progress, selects the next tasks, and decides
% whether to continue, return to an earlier stage, or request human
% input. Human input is particularly important when decisions affect
% contracts, invariants, or verification goals.

% \subsection{Reuse across verifications}
% \label{sec:reuse}

The workflow supports two forms of reuse: verified definitions
and lemmas, and knowledge about how proofs are developed. Before
starting a proof, the VCA is instructed to consult verified proofs
of related algorithms, domain libraries, and examples of previously
resolved proof obligations. 
% N.B. The mapping is not properly maintained by the workflow.
%A mapping from algorithm families to relevant proofs and library modules makes these resources easier to find.

\paragraph{Domain libraries.}
Shared libraries group definitions, predicates, and lemmas into domain-specific modules for both Dafny and Python. General modules provide the mathematical definitions used in specifications, while specialised modules contain results for particular algorithm families, such as low-link and spanning-tree modules. Results established during one verification can be added to these libraries and reused in later proofs. The Python domain library mirrors the Dafny definitions, allowing tests and conjecture checkers to use the same predicates as the formal contracts.

\paragraph{Reuse of proof designs and experience.}
Reuse also extends beyond verified code. Status records, lesson files,
and a repository-wide catalogue preserve proof designs, verification
strategies, and approaches that did not work. These provide useful
starting points for later proofs, although new algorithms still
require domain-specific reasoning. Techniques can also transfer
between different domains. For example, a context-reduction
technique developed during the Lengauer--Tarjan verification of
graph algorithms was later reused in branching bisimulation proofs
in the LTS domain. This shows how experience from one verification
can benefit not only related algorithms but also quite different
verification efforts.

%%%%%%%%%%%%%%%%%%%%%%%%%%%%%%%%%%%%%%%%%%%%%%%%%%%%%%%

\section{Evaluation}
\label{sec:evaluation}

%We evaluate the workflow on a collection of algorithms from
%graph theory, labelled transition systems (LTSs), and computational geometry. 
The evaluation has two parts. First,  we list results for a sizable corpus of algorithms in different domains. Second, we discuss selected algorithms that illustrate challenges encountered during verification and how the workflow addressed them.

\subsection{Verification Corpus}

The verification corpus consists of a collection of 37 verified algorithms from
graph theory, labelled transition systems (LTSs), and computational geometry.  
Table~\ref{tab:verification-results} summarizes verification results for these algorithms. 
The table characterises the size and verification time of the
resulting artefacts, rather than comparing algorithm complexity.
Verification time depends on proof structure, hardware, verifier
configuration, and the amount of proof engineering required.
The \emph{Days} column gives a rough indication of development
time, measured from the first commit of the algorithm's
specification to the first version that passes verification.
All verification runs were performed on a desktop PC with a
13th Gen Intel Core i7-13700 processor (16 cores) and 128\,GB
of RAM. We used Dafny 4.10.0 with its bundled version of Z3.
Each verification run was limited to six cores and a total
memory usage of 12\,GB. Together, the standalone verification files contain more than 74,000 lines of Dafny code, over 2,300 lemmas, and
nearly 5,000 declarations.

The corpus includes several
challenging verification targets, notably Edmonds' blossom
algorithm, Lengauer--Tarjan dominators, and the Hopcroft--Tarjan
triconnectivity test. The n-dimensional Quickhull algorithm is 
also verified with the restriction that the input points are in general position.
 To our knowledge, for both the Hopcroft--Tarjan triconnectivity test
 (based on the version corrected by Gutwenger and Mutzel~\cite{GutwengerM00}),
 and the n-dimensional Quickhull algorithm no previous
machine-checked verification has been reported.

\begin{table}[thbp]
\centering
\caption{Dafny verification of the standalone files of
\texttt{graph}, \texttt{lts}, and \texttt{geometry}.
All measurements use Dafny 4.10.0 with the
\texttt{--cores} option set to \texttt{6},
and a 12\,GB memory limit (see Section~\ref{sec:evaluation}
for the hardware configuration). The \textit{Days} column
shows the elapsed calendar days between the commit of the
algorithm pseudocode and its first successful verification.
These values provide only a rough indication of development
time, not actual time spent working on the algorithm. They
may include extended periods during which verification was
stalled pending new proof techniques, as was the case for
\texttt{biconnected-components}.}

% All measurements use Dafny 4.10.0, \texttt{--cores 6},
% and a 12\,GB memory limit (see Section~\ref{sec:evaluation}
% for the hardware configuration). Verified 2026-11-09.}

\label{tab:verification-results}
\small
\begin{tabular}{l r r r r r}
\toprule
Algorithm & LoC & Lemmas & Declarations & Time (s) & Days \\
\midrule
\multicolumn{6}{l}{\emph{Graph algorithms}} \\
\texttt{a-star} \cite{a-star}                              & 288 & 7 & 17 & 3.3 & 1 \\
\texttt{articulation-points} \cite{tarjan}                 & 3724 & 63 & 127 & 79.0 & 96 \\
\texttt{bellman-ford} \cite{bellman-ford}                  & 570 & 13 & 38 & 7.8 & 1 \\
\texttt{biconnected-components} \cite{biconnected-components} & 6039 & 134 & 273 & 74.1 & 123 \\
\texttt{boruvka} \cite{boruvka}                            & 3008 & 82 & 150 & 19.8 & 108 \\
\texttt{bridges} \cite{tarjan}                             & 3560 & 59 & 119 & 65.9 & 91 \\
\texttt{bron-kerbosch} \cite{bron-kerbosch}                & 783 & 32 & 61 & 6.8 & 1 \\
\texttt{connected-components} \cite{connected-components}  & 274 & 7 & 17 & 1.7 & 55 \\
\texttt{dijkstra} \cite{dijkstra}                          & 456 & 11 & 27 & 8.6 & 1 \\
\texttt{dynamic-connectivity} \cite{dynamic-connectivity}  & 1826 & 47 & 87 & 12.0 & 1 \\
\texttt{edmonds-blossom} \cite{edmonds-blossom}            & 3285 & 82 & 175 & 34.4 & 1 \\
\texttt{edmonds-karp} \cite{edmonds-karp}                  & 853 & 25 & 77 & 4.6 & 30 \\
\texttt{fleury} \cite{fleury}                              & 1422 & 52 & 110 & 3.8 & 3 \\
\texttt{floyd-warshall} \cite{floyd-warshall}              & 768 & 21 & 58 & 9.7 & 1 \\
\texttt{ford-fulkerson} \cite{ford-fulkerson}              & 832 & 23 & 73 & 8.3 & 30 \\
\texttt{hierholzer} \cite{hierholzer}                      & 851 & 32 & 76 & 22.0 & 1 \\
\texttt{hopcroft-karp} \cite{hopcroft-karp}                & 1502 & 47 & 102 & 14.6 & 1 \\
\texttt{johnson} \cite{johnson}                            & 1257 & 39 & 102 & 17.2 & 38 \\
\texttt{kosaraju} \cite{kosaraju}                          & 723 & 26 & 52 & 2.7 & 38 \\
\texttt{kruskal} \cite{kruskal}                            & 2300 & 65 & 121 & 19.8 & 99 \\
\texttt{kuhn} \cite{kuhn}                                  & 817 & 23 & 46 & 7.0 & 30 \\
\texttt{lengauer-tarjan} \cite{lengauer-tarjan}            & 3812 & 126 & 288 & 122.1 & 1 \\
\texttt{moore} \cite{moore}                                & 557 & 22 & 60 & 3.5 & 1 \\
\texttt{prim} \cite{prim}                                  & 1361 & 34 & 63 & 7.0 & 84 \\
\texttt{push-relabel} \cite{push-relabel}                  & 1336 & 49 & 131 & 17.3 & 66 \\
\texttt{schorr-waite} \cite{schorr-waite}                  & 380 & 19 & 42 & 25.6 & 2 \\
\texttt{suurballe} \cite{suurballe}                        & 1875 & 64 & 142 & 33.5 & 13 \\
\texttt{tarjan} \cite{tarjan}                              & 1599 & 41 & 101 & 17.3 & 1 \\
\texttt{triconnectivity-test} \cite{triconnectivity-test}  & 9731 & 370 & 794 & 226.4 & 14 \\
\texttt{ullmann} \cite{ullmann}                            & 662 & 22 & 58 & 4.3 & 1 \\
\texttt{welsh-powell} \cite{welsh-powell}                  & 805 & 27 & 64 & 3.8 & 1 \\
\midrule
\multicolumn{6}{l}{\emph{LTS algorithms}} \\
\texttt{drwb} \cite{drwb}                                  & 2695 & 109 & 227 & 25.5 & 1 \\
\texttt{failures-antichain-check} \cite{failures-antichain-check} & 1125 & 43 & 94 & 21.8 & 1 \\
\texttt{groote-jansen} \cite{groote-jansen}                & 4109 & 177 & 246 & 47.6 & 1 \\
\texttt{groote-vaandrager} \cite{groote-vaandrager}        & 1080 & 40 & 54 & 17.4 & 1 \\
\texttt{jgkw} \cite{jgkw}                                  & 3137 & 133 & 173 & 54.8 & 1 \\
\midrule
\multicolumn{6}{l}{\emph{Geometry algorithms}} \\
\texttt{quickhull-nd} \cite{quickhull-nd}                  & 5137 & 205 & 477 & 170.4 & 9 \\
\midrule
Total                                                             & 74539 & 2371 & 4922 & 1221.7 &  \\
\bottomrule
\end{tabular}
\end{table}

\subsection{Case Study}

We discuss four algorithms that illustrate different aspects
of the workflow: invariant discovery in triconnectivity test,
large-scale proof restructuring in Lengauer--Tarjan, largely
autonomous verification of Edmonds' blossom, and the transfer
of proof-engineering techniques to LTS algorithms. The selected
algorithms involve substantial mathematical reasoning and
present different challenges for formal verification.

\paragraph{Triconnectivity test.}
The Hopcroft--Tarjan triconnectivity test is our largest and
most demanding algorithm. Its correctness contract is expressed
in terms of graph separation, while the algorithm reasons about
depth-first search trees, low points, and separation pairs.
Connecting these two views required substantial invariant
development. In particular, the completeness invariant for
the path-search phase went through many revisions before it
became both correct and inductive. The conjecture-testing
infrastructure helped identify missing conditions and improve
candidate invariants.

After the invariants had been established, proof completion
still required considerable proof engineering. Large verification
contexts and quantifier interactions caused solver problems
that were addressed through opaque predicates, explicit lemmas,
and assertion isolation. This case illustrates the role of
empirical invariant validation and solver-guided proof
optimisation in a large verification effort.

\paragraph{Lengauer--Tarjan.}
The Lengauer--Tarjan dominator algorithm
illustrates the importance of proof restructuring. The main
problem was an excessively large verification context, which
caused repeated solver timeouts. This is addressed by hiding
large contracts and invariants behind opaque predicates and
exposing only the facts needed at specific proof boundaries.
Explicit lemmas were then introduced to control how these
facts were made available.

The combined restructuring reduced the number of obligations
in the main verification routine from 644 to 271, eliminated
36 timeouts, and reduced Z3 resource consumption from
approximately 2.1 billion to 144 million resource units.
The improvement illustrates how measurements of solver
behaviour can guide changes to proof structure and turn an
otherwise impractical verification into a successful one.

\paragraph{Edmonds' blossom.}
Edmonds' blossom algorithm provides a different test of the
workflow: an end-to-end verification of a complex algorithm
with limited human involvement. The algorithm combines
alternating forests, blossom contraction, and path lifting,
requiring contracts and supporting lemmas that connect these
different parts.

The verification started from a published algorithm description
and was informed by the detailed Isabelle verification by
Abdulaziz et al.~\cite{AbdulazizM26}. The human role was
limited to selecting the problem and monitoring progress.
The VCA developed the contracts, invariants, and proof artefacts,
and addressed verification failures. After assertion isolation
was applied to a small number of declarations, the complete
implementation passed Dafny's native verification. The
verification covered more than 1,100 proof obligations across the
domain library for graph matching and the algorithm's verification. This case
demonstrates that the workflow can handle a substantial
verification task with little direct human guidance.

\paragraph{Groote--Jansen and JGKW.}
With these two algorithms we examine whether the workflow's
proof-engineering techniques also apply outside graph algorithms.
Both concern LTS algorithms and required substantial proof
engineering. In Groote--Jansen, context restructuring,
supporting lemmas, and assertion isolation were needed to
resolve repeated timeouts. JGKW presented similar problems,
which were overcome through resource-guided restructuring
and selective assertion isolation.

These results provide evidence that techniques developed
during graph-algorithm verification can also help with
algorithms from a different domain. Although two examples
are not sufficient to establish general transferability,
they show that the workflow's proof-engineering techniques
are not limited to graph algorithms.

%%%%%%%%%%%%%%%%%%%%%%%%%%%%%%%%%%%%%%%%%%%%%%%%%%%%%%%

\section{Discussion}
\label{sec:results}

\subsection{Repository-Wide Results}

The workflow has been applied to algorithms from several domains,
including graph theory, labelled transition systems, game-tree search,
sorting, and geometry. The resulting Dafny proofs range from a few
hundred to almost ten thousand lines of code. Several targets,
including Edmonds' blossom algorithm, the
Lengauer--Tarjan dominator algorithm, and the Hopcroft--Tarjan
triconnectivity test, are particularly challenging because of
their complex invariants and global correctness requirements.

All algorithms we attempted to verify, except Hopcroft--Tarjan's tri\-con\-nected-com\-ponents algorithm have been fully verified. Its ongoing verification has reached 16,658 lines of Dafny code and 460 lemmas. This suggests that the practical limits of Dafny verification have not yet been reached, as new proof-engineering techniques continue to enable progress on previously stalled problems.

The verification corpus was developed over several months.
Many of the proofs would represent substantial verification
efforts in their own right. Although we have no controlled
comparison with conventional verification, the results show
that a structured LLM-assisted workflow can produce machine-checked
proofs of challenging algorithms across a range of domains.

\subsection{Cross-Case Findings}
\label{sec:cross-case}

Three findings recurred across the verifications.
First, the main difficulties were often related to proof engineering
rather than to the underlying mathematics. Large contexts visible
to the solver, interactions between quantified predicates, and
groups of proof obligations that were costly to verify together
often caused problems, even when the required mathematical facts
were already available. In such cases, restructuring the proof
context was more effective than adding further facts.

Second, measurements were important for diagnosing verification
problems. Solver resource consumption, timeouts, and the structure
of proof obligations helped distinguish incorrect statements,
missing proof facts, and performance problems. This made it
possible to choose targeted solutions instead of relying on
trial and error.

Third, proof-engineering techniques could often be reused across
verifications. Opaque invariant bundles, assertion
isolation, proof decomposition, and specialised library lemmas
were applied to different algorithms. Notably, assertion isolation
was especially needed in almost all complex verifications.
These findings suggest that a
central challenge in LLM-assisted verification is not simply
generating proof steps, but organising proofs so that automated
reasoning tools can handle them effectively.

\subsection{Evolution of the Workflow and Human Oversight}
\label{sec:evolution}

The workflow developed in response to problems encountered during
verification. Several algorithms, including \texttt{boruvka}, \texttt{bridges},
and \texttt{biconnected-components}, initially stalled for weeks despite
considerable proof-development effort. Progress resumed after
diagnostic and optimisation capabilities were added. Measurements
of solver resource usage helped identify performance problems and
guided changes to the proofs. Successful techniques were then
recorded for use in later verifications.

The human role changed substantially as the workflow matured.
In the early stages, the human actively contributed to proof
development by proposing and discussing proof techniques,
suggesting a change in the proof strategy when verification became
stuck (often with help from another agent), and occasionally
making major changes to the workflow. In the mature workflow,
the VCA can carry out most of the reasoning autonomously.
Human involvement is now mainly limited to monitoring progress,
making occasional decisions about the direction of the proof
effort, and intervening when the workflow fails to make
progress. For example, during the ongoing verification of
triconnected components, the conjecture-generation step became
stuck in a loop without making significant progress. After
discussion, the idea of using a proof dependency graph emerged.
The agent then generated the graph and used it to guide a
bottom-up proof effort, which proved considerably more
successful. Occasionally, the human needs to extend the tool support,
e.g. when a new pseudocode language construct  is needed. Thus, the
human role has shifted from actively guiding proof development
to primarily supervising the process and only intervening when
necessary.

\subsection{Trustworthiness and Limitations of the Evidence}
\label{sec:trustworthiness}

The correctness claims depend on the reconstructed pseudocode and
formal contract accurately describing the intended algorithm
(Steps~1--2). Ensuring their correctness is simply the responsibility
of the human. 
Their review may nevertheless fail to detect errors or omissions, which may compromise the entire verification. 
The subsequent generation of Dafny code from the pseudocode and
contract (Step~3) may also introduce errors, which must be
detected through testing and verification. On the other hand, guardrails prevent
changes that could weaken the locked verification target.
Finally, the correctness evidence depends on the reliability
of the testing and verification procedures, in particular the
Dafny verification of the complete file in Step~9. As in
any Dafny verification, the correctness claims remain conditional
on the soundness of Dafny.
% The correctness claims depend on the reconstructed pseudocode and
% formal contract accurately describing the intended algorithm.
% Ensuring their correctness is the responsibility of the human.
% Guardrails prevent later changes that could weaken the verification target.
% Dafny's verification and audit mechanisms check that all proof
% obligations are discharged and that assumptions, axioms, and
% bodyless lemmas are not used. As in any Dafny verification,
% the correctness claims remain conditional on the soundness of
% Dafny and its SMT solver Z3.

% The evaluation has several limitations. The algorithms were
% selected by the authors rather than taken from a predefined
% benchmark, and no controlled comparison with human-led
% verification was performed. Some algorithms may also be
% represented in the training data of contemporary LLMs.
% Finally, the workflow changed throughout the project, making
% verification campaigns from different stages difficult to
% compare directly. The results therefore demonstrate the
% effectiveness of the final workflow on a collection of challenging
% algorithms, but do not establish how it would perform on arbitrary
% software or how much each individual component contributes.

The evaluation has several limitations. We selected well-known, published algorithms that are represented in the training data of contemporary LLMs, although the extent to which they are represented is unknown. This is relevant when interpreting the results, since prior exposure to an algorithm may influence the effort required to reconstruct and verify it. Although the workflow successfully produced verified Dafny implementations for all algorithms in the collection, this does not establish how well it generalizes to unpublished or previously unseen algorithms, or to arbitrary source code rather than published algorithm descriptions. No controlled comparison with human-led verification was performed. Finally, the evaluation does not isolate the contribution of each individual component of the workflow.

\section{Conclusion and Future Work}
\label{sec:conclusions}

We have presented a workflow for LLM-assisted verification
of algorithms in Dafny. It combines contract development,
implementation and testing, empirical invariant validation,
iterative proof construction, and resource-guided diagnosis and
optimisation. Rather than introducing a new verifier or proof
technique, the workflow brings together LLMs, verification tools,
empirical checks, and human supervision in a systematic approach to
formal verification.

We applied the workflow to challenging algorithms, including
Edmonds' blossom algorithm, the Lengauer--Tarjan
dominator algorithm, and the Hopcroft--Tarjan triconnectivity test.
The results show that substantial machine-checked proofs can be
developed through a structured LLM-assisted process. A key finding
is that the main difficulties often do not lie in the underlying
mathematics, but in proof engineering: controlling verification
contexts, structuring proof obligations, and dealing with solver
performance. Resource-guided diagnosis and proof optimisation
were essential for overcoming these difficulties. As the workflow
matured, the VCA became capable of carrying out most proof reasoning
autonomously, while the human role shifted towards monitoring
progress, making occasional decisions, and intervening when
necessary. The resulting correctness claims remain conditional
on the quality of the specifications and the soundness of Dafny
and its underlying SMT solver, Z3.

Several directions for future work remain. First, the workflow
should be evaluated on algorithms and domains less likely to be
represented in LLM training data, and on existing software
implementations rather than reconstructed pseudocode. It would
also be interesting to apply the same approach in other verification
environments, such as Lean, Isabelle, and Coq.

Second, the repository provides an opportunity to study verification
processes in greater detail. Its records of proof attempts, solver
resource usage, and proof-engineering decisions could help identify
common causes of verification failures and evaluate the effectiveness
of individual techniques and workflow components, particularly
empirical invariant validation and resource-guided optimisation.
Further applications to challenging algorithms will also help
identify remaining bottlenecks and improve the workflow.

Finally, successful verification does not necessarily produce
readable proofs. An important direction is therefore to improve
proof readability and maintainability alongside successful
verification. The available proof histories and transcripts provide
a basis for studying how LLM-generated proofs develop and how
their resulting artifacts can be made easier for humans to
understand and maintain.

% ------------------------------------------------------------
% Bibliography
% ------------------------------------------------------------
\subsubsection{Data-Availability Statement}
The source code, Dafny verification models, and instructions for running the
experiments supporting the results in this paper are available at \url{https://10x10.org/dafny}.

\subsubsection{\discintname}
The authors declare no competing interests relevant to the content of this
article.

\subsubsection{Declaration of AI-Assisted Writing}
During the preparation of this manuscript, the authors used ChatGPT (OpenAI)
to improve wording and language clarity. The authors reviewed and revised
the suggested formulations and take full responsibility for the final
content of the manuscript.

\bibliographystyle{splncs04}
\bibliography{main}

\end{document}

%% file: workflow.tex
\begin{tikzpicture}[
    >=Latex,
    node distance=22mm and 22mm,
    phase/.style={
        rectangle,
        rounded corners=2mm,
        draw=black!70,
        thick,
        fill=blue!8,
        minimum width=34mm,
        minimum height=10mm,
        align=center
    },
    review/.style={
        diamond,
        draw=black!70,
        thick,
        fill=pink!25,
        aspect=1.3,
        inner sep=1.5pt,
        align=center,
        font=\small,
        minimum width=2.5cm
    },
    flow/.style={->,thick,black!75},
    feedback/.style={->,very thick,green!55!black},
    publication/.pic={
        \draw[thick, fill=blue!8]
            (0,0) -- (1.5,0) -- (1.5,1.7) -- (1.1,2.1)
            -- (0,2.1) -- cycle;
        \draw[thick] (1.1,2.1) -- (1.1,1.7) -- (1.5,1.7);
        \node[align=center,font=\scriptsize] at (0.75,0.85)
            {Research\\article};
    }
]

% Top row
\node[phase,fill=yellow!25] (spec) {Specification\\ (Steps 1-2)};

% Publication input below spec
\node[below= 15mm of spec, minimum width=11.25mm, minimum height=15.75mm,
      inner sep=0pt, outer sep=0pt] (pub) {};
\pic[scale=0.75] at (pub.south west) {publication};
\draw[flow] (pub.north) -- (spec.south);

\node[review,right=of spec] (reviewA) {Human\\Review};
\node[phase,fill=blue!25,right=of reviewA] (testing) {Code Generation \& Contract Testing\\ (Steps 3-4)};
	\node[phase,fill=green!15,right=of testing] (proofdesign) {Proof Design \& Validation\\(Steps 5--6)};

% Bottom row (right to left)
\node[phase,fill=green!15,below=of proofdesign] (proofcomp) {Proof Construction\\(Step 7)};
\node[phase,left=of proofcomp, below=of testing, fill=green!15] (verify) {Proof Verification \& Optimization\\(Steps 8--9)};
\node[phase, anchor=center] (export) at (reviewA.center |- verify.center){Export\\(Step 10)};

% Main flow
\draw[flow] (spec) -- 
    node[midway,above,font=\small, align=center]{\emph{concept}}
    node[midway,below,font=\small, align=center]{contract \& \\pseudocode}
    (reviewA);

% revision loop:
\coordinate (route1) at ($(spec)!0.5!(reviewA)+(0,-1.25cm)$);
\coordinate (route2) at ($(spec.south)!0.5!(spec.south east)$);
  \draw[flow] (reviewA.south)
    -- (reviewA.south |- route1)
    node[midway,below=2mm,font=\small] {revision}
    -- (route2 |- route1)
    -- (route2);

  \draw[flow,-] (testing.south)
    -- (testing.south) |- (route1);

\draw[flow]
    (reviewA) -- 
    node[midway,above,font=\small,align=center]{\emph{locked}} 
    node[midway,below,font=\small,align=center]{contract \& \\pseudocode} 
    (testing);
%\draw[flow] (reviewA) -- (testing);
\draw[flow] (testing) -- (proofdesign);
\draw[feedback] (proofdesign) -- node[midway,right,font=\small,align=center]{proof plan} (proofcomp);
\draw[feedback] (proofcomp) -- node[midway,below,font=\small,align=center]{proof draft} (verify);
\draw[flow] (verify) -- (export);

% human supervision
\coordinate (triCenter) at (barycentric cs:proofdesign=1,proofcomp=1,verify=1);
\node[cloud, draw=black!70, fill=pink!25, aspect=2.0, scale=0.9, align=center,
      inner sep=2pt] (oversight)
      at ($(triCenter)+(0,5mm)$)
      {Human\\supervision};

\draw[dashed, thick] (oversight) -- (proofdesign);
\draw[dashed, thick] (oversight) -- (proofcomp);
\draw[dashed, thick] (oversight) -- (verify);

% Feedback loop
\draw[feedback] (verify.north)
    to node[above,font=\small,sloped] {proof redesign} ($(proofdesign.west)!0.5!(proofdesign.south west)$);

% self loop
\coordinate (loopstart) at  ($(verify.north west)!0.25!(verify.north east)$);
\draw[flow,green!55!black]
  (loopstart)
  .. controls ($(loopstart)+(-10mm,9mm)$)
             and ($(loopstart)+(10mm,9mm)$) ..
  node[midway,above,font=\small,green!55!black] {subagents}
  ($(loopstart)+(2mm,0)$);

% repository
    \coordinate (repoSW) at
  ([xshift=-2mm,yshift=-6mm]current bounding box.south west);
\coordinate (repoNE) at
  ([xshift=2mm,yshift=4mm]current bounding box.north east);

\draw[dashed,thick] (repoSW) rectangle (repoNE);

\node[anchor=south west,fill=white,inner sep=1pt]
  at ([xshift=2mm,yshift=0.5mm]repoSW) {\textbf{repository}};

\end{tikzpicture}